\documentstyle[12pt,fleqn]{article}
\def\beq{\begin{equation}}
\def\eeq{\end{equation}}
\def\sT{{\sf T}}
\def\sA{{\sf A}}
\def\1{\mbox{\small1\hskip-0.35em\normalsize1}}
\def\6{\langle }
\def\9{\rangle }

\begin{document}

\renewcommand{\thefootnote}{\fnsymbol{footnote}}

\vspace*{10mm}
\begin{center}
{\large {\bf Non-linear operations in\\[1mm] 
quantum information theory}}\\[15mm]

 Daniel R. Terno\footnote{Electronic address:
terno@physics.technion.ac.il} \\[8mm]
{\sl Department of Physics, Technion---Israel Institute of Technology,
32\,000 Haifa, Israel}\\[15mm]
\noindent{\bf Abstract}\end{center}
Quantum information theory is used to analyze various non-linear operations
on quantum states. The universal disentanglement machine is shown to be
impossible, and partial (negative) results are obtained in the state-dependent
 case. The efficiency of the transformation of
non-orthogonal states into orthogonal ones is discussed.
\vfil
\vfil

\newpage

\noindent 
The rules of a quantum mechanics make certain processes
impossible. Non-orthogonal quantum states connot be cloned~[1].
This is one of the fundamental theorems of the quantum information theory. On
other hand, there are explicit constructions of quantum circuits that 
would perform many interesting transformations of quantum states~[2, 3],
provided that certain devices, like quantum XOR gates~[4], can be actually 
built. Between these two extremes there is a gray area of
operations that are not obviously forbidden by the elementary laws of quantum 
mechanics, but may be ruled out
by more careful considerations. The most interesting of them
 are non-linear operations, 
and their analysis from the point of view of quantum information theory~[5, 6]
is the subject of this paper. 

The basic procedures of quantum mechanics --- unitary transformations 
and projections --- are linear. However, non-linear 
operations with quantum states are common, even if they are not always 
regarded as such.
Non-linearity naturally enters via selective operations, namely
those that involve a filtering at one of their steps. These operations
 have a finite probability to fail and may look quite exotic, e.g., 
the transformation~[2]
\beq
\rho^{{\rm in}}=\left( \begin{array}{cc}
\rho_{11} & \rho_{12} \\
\rho_{21} & \rho_{22}
\end{array} \right) \rightarrow \rho^{{\rm out}}=
\left( \begin{array}{cc}
\rho_{11}^2 & \rho_{12}^2 \\
\rho_{21}^2 & \rho_{22}^2
\end{array} \right).\label{right}
\eeq
Nevertheless, the possibility of failure is not a necessary condition of
non-linearity. An operation may be non-selective, i.e. it is always be 
completed successfully, but there may be a demand for certain inputs to 
be transformed
into certain outputs, while the processing of other input states remains 
unspecified. For example, if we want to process the output of a two-state 
cloning
machine~[3], the transformation of the two resulting pure entangled states 
(labelled 1 or 2) into the direct product of their reduced density matrices,
\beq
\rho_{1,2} \rightarrow {\rm Tr_A}\rho_{1,2}\otimes  {\rm Tr_B}\rho_{1,2}
\label{dis}
\eeq
would be useful. What happens to all other states is irrelevant and 
unspecified; everything that makes this process work will be accepted. 
If we are interested only in these two states, the transformation looks 
`non-linear'.  

The most general physical operation
\beq
\rho\rightarrow {\rho'}=\frac{{\sf T}\rho}{{\rm Tr}\sT \rho} \label{er}
\eeq
is represented by a superoperator 
$\sT: {\cal B}({\cal H})_1 \rightarrow {\cal B}({\cal H})_1 $, 
which is (i) positive, (ii) satisfies $\sT^{\dag} \sT\leq {\sf 1} $,
and (iii) is completely positive~[7, 8].  The validity of last condition~[9] 
has recently become a hot issue in numerous discussions, but it certainly 
holds for systems which are initially unentangled with their environment.

 According to the `first representation theorem'~[8]
\beq
\sT \rho=\sum_k {\sf A}_k\rho{\sf A}_k^{\dag}, \label{r1}
\eeq  
 where the set of ${\sf A}_k$'s satisfies
\beq
\sum_k {\sf A}_k^{\dag} {\sf A}_k\leq {\sf 1}. \label{rep}
\eeq
Since the probability to get a certain outcome is given by  
$p=\rm Tr \sT\rho$\/,
the transformation in Eq.~(\ref{dis}) is trace preserving, so that
$\sum_k {\sf A}_k^{\dag} {\sf A}_k={\sf 1}$.

When a `quantum black box'~[10] is specified, the superoperator $\sT$ and the
corresponding matrices ${\sf A}_k$ can be inferred from 
the complete set of input and output data. Similarly, if 
we extend Eq.~(\ref{dis}) to a larger set of states
 and then use the approach of Chuang 
and Nielsen~[10], we can confirm or refute different realizations 
of the desired transformation. If the answer is positive the problem is solved.
On the contrary, a negative 
answer tells us nothing. Another realization may still work.

Quantum information theory can discern absolutly impossible 
processes from tentatively possible ones. We look at the desired operation as 
a part of some hypothetical decision scheme, which aims at distinguishing 
between different input 
states~[6, 11, 12], or as a communication channel~[13] with input and output
alphabets given by the left and right hand sides of an expression like 
Eq.~(\ref{dis}).

Using Eqs.~(\ref{er}, \ref{r1}) it is easy to show that any chain of the
 measurements and/or
 transformations of the system, with or without $ancilla$, can be described as
 a single measurement~[14] represented by a positive operator-valued measure 
(POVM). Thus if the proposed procedure allows to 
improve {\em any} of the distinguishability criteria of the input
 states beyond the optimal value, 
it contains some flaw and therefore is not physical.

One of these distinguishability criteria is the probability of error (PE),
defined as follows~[6, 11]: an observer is 
given one of two states $\rho_1$ and $\rho_2$. That state is secretly 
chosen, and since we are interested in the intrinsic difference between
the states, the probilities to pick up either of the states are equal.
 The task of the observer is to decide, after performing anything
 that is allowed by quantum mechanics, which state was given to 
him. The probability
that the observer makes a wrong guess with the best possible decision scheme 
gives a measure of the distinguishability.
Two orthogonal states can be distinguished perfectly, thus giving
${\rm PE=0}$. When the states are identical ${\rm PE}=\frac{1}{2}$. In
general the optimal result is
\beq
{\rm PE}(\rho_1,\rho_2)=\frac{1}{2}-\frac{1}{4}{\rm Tr}|\rho_1-\rho_2|.
\eeq
Any transformation that would give a lower PE certainly violates the laws of
quantum mechanics. Unfortunately, the usefulness of this criterion is limited 
to the case of two states, which is the only one for which a closed solution
 is known.

Another criterion is the accessible information $I(\rho_1,\rho_2)$, 
which is
defined as a maximal mutual information over all possible decision
schemes. For a set of states $\rho_i$ with fixed {\em a priori} 
probabilities $\pi_i$ each possible measurement scheme $X$ 
gives a probability distribution of the results. We calculate the 
mutual information between the probability distributions corresponding 
to different states. Finally the maximum is taken over all possible 
measurements. 
In the case of two inputs the accessible information is given by [5, 6, 12, 13]
\beq
I(\rho_1,\rho_2):=\max_X [I(p(\rho_1,X),p(\rho_2,X))].
\eeq
This definition is naturally extended to more than two states. Accessible
information cannot increase, but there are only a handful of  cases 
where it is explicitely known, in particular two pure states and two 
spin-$\frac{1}{2}$ states~[15]. 

When $I$ is unknown we can use different  unequalities that relate 
the accessible information to other distinguishability criteria~[6, 12]. 
The most useful
of them involves a new notion, which is called 
entropy defect or relative entropy~[5] and is given by 
\beq
\Delta S(\rho_1,\ldots,\rho_n)=S(\bar{\rho})-\sum_i\pi_i S(\rho_i),
\eeq
where 
\beq
S(\rho)=-{\rm Tr}~ \rho \log \rho,
\eeq
is the von Neumann entropy, $\bar{\rho}=\sum_i \pi_i \rho_i$, and $\pi_i$ is 
the $a~ priori$ probability distribution. In order to measure the intrinsic
difference between two of the states, we set $\pi_1=\pi_2=\frac{1}{2}$, as in 
the definition of PE. 
The Levitin-Holevo inequality states that the entropy deffect is an upper 
bound of the accessible information~[5]
\beq
I(\rho_1,\rho_2)\leq \Delta S(\rho_1,\rho_2).
\eeq
Moreover, the entropy defect does not increase under trace preserving
completely positive maps~[13],
\beq
\Delta S(\sT\rho_1,\sT\rho_2)\leq \Delta S(\rho_1,\rho_2).
\eeq
It is obvious that the compliance with known information bounds is only a
 necessary condition that the proposed transformation must satisfy. However, 
it is a powerfull tool, as the following examples illustrate. 

Since the exact cloning of quantum states, $ \rho\rightarrow\rho\otimes\rho$ is
impossible~[1], thelet us consider the best approximate cloners. 
The optimal two-state
cloning machine~[3] is a device which has one of the two 
possible pure spin-$\frac{1}{2}$ states as the input and produces an entangled
pair on the output. Two identical reduced density matrices are close to the 
cloned state with exceptionally high fidelty,
 $F={\rm Tr}[({\rm Tr}_A \rho^{{\rm out}}_i)|\psi^{{\rm in}}_i\9
\6\psi^{{\rm in}}_i|]>0.985$. Although the fidelity of the reduced density
matrices is very high, it will be shown that it is impossible to separate
the output according to the prescription of Eq.~(\ref{dis}). If the
two input states are parametrized as
\begin{eqnarray}
|{\rm u}\9  =  \cos\theta|0\9+\sin\theta|1\9, \nonumber \\
|{\rm v}\9  =  \sin\theta|0\9+\cos\theta|1\9,
\end{eqnarray}
the output states are
\begin{eqnarray}
|{\rm u'} \9  =  (a \cos\theta+c \sin\theta )|00\9
 +b(\cos\theta+\sin\theta)(|01\9+|10\9)+
(c\cos\theta+a\sin\theta)|11\9, \nonumber \\
|{\rm v'}\9  =  (c \cos\theta+a \sin\theta )|00\9  
 +b(\cos\theta+\sin\theta)(|01\9+|10\9)+
(a\cos\theta+c\sin\theta)|11\9, \label{new}
\end{eqnarray}
where $a(\theta)$, $b(\theta)$ and $c(\theta)$ are given in the Appendix.

Let us suppose that there is a disentanglement procedure which is described 
by Eq.~(\ref{dis}). Then we can include the cloning machine and the 
`disentangler' into Helstroms's decision scheme. 
Instead of performing the optimal 
measurement on the output states $\rho^{\rm out}_i$ (which gives the same PE
 as the measurement on $\rho^{\rm in}_i$, since they are unitarily related)
we perform it on the direct product 
of its disentangled copies,
 ${\rm Tr_A} \rho^{\rm out}_i\otimes {\rm Tr_B} \rho^{\rm out}_i$.
In both cases the analytical expressions for PE can be found explicitely. They
 are given in the Appendix and their graphs are plotted on Fig.~1. We see 
that the disentanglement of the  copies decreases PE and, 
as a result, this process is impossible. 
Closer look on the construction of the optimal measurement reveals that the 
proposed transformation is realized by an operator which is not $positive$
and, consequently, can represent no physical process.

Before looking at the disentanglement of more general states, it is easy
to see why the universal disentanglement machine is impossible.
A transformation
\beq
\rho\rightarrow {\rm Tr_A}\rho\otimes{\rm Tr_B}\rho,
\eeq
 for all input states is essentially 
non-linear. Thus it cannot correspond to any physical process.
 
 Now, let us again consider the states 
\begin{eqnarray}
|{\rm u}\9=a|00\9+b(|01\9+|10\9)+c|11\9,\\ \nonumber
|{\rm v}\9=c|00\9+b(|01\9+|10\9)+a|11\9,
\end{eqnarray}
 where the coefficients $a$, $b$ and $c$ are arbitrary
real numbers subject only to the normalization, $a^2+2 b^2+c^2=1$.
They can be parametrized by spherical coordinates as
\beq
a=\sin\vartheta \cos\varphi \\ b=\sin\vartheta \sin\varphi/\sqrt{2} \\
c=\cos\vartheta .
\eeq
It is possible to derive analytical expressions for PE and $\rm PE_d$, which
 are given in the Appendix.
 ${\rm PE}(\vartheta,\varphi)$ is the actual optimal
result, while the decision process which leads to  
${\rm PE_d}(\vartheta,\varphi)$ includes the hypothetical disentanglement 
procedure. 
Obviously, the regions of the $(\vartheta, \varphi)$ plane 
where ${\rm PE}<{\rm PE_d}$ are forbidden, i.e., the disentanglement procedure
cannot be realized. Fig. 2 shows these areas together with the line that 
corresponds to the disentanglement of the output of the optimal cloner, which 
lies in one of the forbidden domains.

Another bound can be obtained by using the entropy defect.
The explicit expressions 
for $\Delta S$ and $\Delta S_{\rm d}$ will
 not be given here, because they are too cumbersome. Proceeding exactly as in 
the previous case, we look for the regions
where $\Delta S\leq \Delta S_{\rm d}$, i.e. where the disentanglement
 (\ref{dis})
is impossible.These domains are presented on Fig.~3. It is instructive
to compare it with Fig.~2. The boundary of the regions forbidden  
by PE coincides with some parts of the boundary obtained by 
$\Delta S$. However, the differences are clear.
This partial agreement  requires further investigation and may bring some
new insightes on the relationship between different distinguishability 
criteria.

In the examples that we considered above, these criteria give identical 
predictions for the states 
with equal degree of entanglement. For pure states the degree of entanglement
is measured by von Neumann entropy of either 
subsystem~[16],
\beq 
E(\rho)=S({\rm Tr_{A}}\rho)=S({\rm Tr_{B}}\rho).
\eeq
The eigenvalues of either of $\rm Tr_A |u\9\6u|$ or $\rm Tr_A|v\9\6v|$ are
\beq
\lambda_{\pm}=\frac{1}{2}\left( 1\pm(a+c)\sqrt{1+2 b^2-2ac}\right),
\eeq
These eigenvalues (and, as a result, the degrees of the entanglement of
the corresponding states) are unchanged under the transformation
$\vartheta\rightarrow \pi-\vartheta,\/\varphi\rightarrow\pi-\varphi$. It
is also a symmetry of Eqs.~(A.6-A.9) and the expressions for $\Delta S$.

The existence of a general `no-disentanglement theorem' remains an open issue.
 However, it seems that the disentanglement is impossible in general and the 
way to prove
it is similar to the proof of the `no-broadcasting theorem'~[6, 17].

A more intricate question~[17] is whether there is a transformation that takes
an entangled state into a separable one~[18] and preserves the reduced density
operators,
\beq
\rho\rightarrow \tilde\rho=\sum_i w_i \rho^{\rm A}_i\otimes \rho^{\rm B}_i, 
\\ \rm Tr_A \tilde\rho=Tr_A \rho,\\ \rm Tr_B \tilde\rho=Tr_B \rho. 
\eeq
This is still to be answered.

As another example, we analyze the information approach to the transformation 
of pure non-orthogonal states into orthogonal ones. Recently~[2] a quantum
circuit that
does this operation for spin-$\frac{1}{2}$ states was proposed. Since 
non-orthogonal 
quantum states cannot be distinguished with certainty, such a transformation
has only a limited probability of success. It consists in applying a 
XOR gate~[4] to the pair of identically prepared particles in either of the
states and measuring the spin of the second particle in $z$ direction. If the
spin is `down' the transformation succeeds. In this case the 
outputs are
 $|\phi^{\rm out}\9_1\otimes|\downarrow\9$ or 
$|\phi^{\rm out}_2\9\otimes|\downarrow\9$; and we have 
$\6\phi^{\rm out}_2|\phi^{\rm out}_1\9=0$ even if initially  there was an 
overlap between the states.

Leaving the explicit calculations aside, let us look at the bounds on this 
operation. Orthogonal states can be, in principle, distinguished unambiguously.
Thus it is possible to use this procedure as part of an error-free 
scheme of discrimination between two non-orthogonal states~[2]. The exact
 solution of this problem ~[19] is given by a POVM with two otputs 
which correspond to the
unambiguous results, $\sA_1$ and $\sA_2$, and an output which is a failure 
of the measurement, $\sA_? ={\sf 1}-\sA_2-\sA_1$. The optimal POVM
 is explicitely known and the probability to have a definite 
answer is
\beq
P=1-|\6\phi_1|\phi_2\9|.
\eeq
Since the direct products of two copies of the same states have an overlap
equal to $|\6\phi_1|\phi_2\9|^2$ the probability of success of
 any transformation of two non-orthogonal states into orthogonal ones 
is bounded by
\beq
P({\rm success})\leq 1-|\6\phi^{\rm in}_1|\phi^{\rm in}_2\9|^2 \label{maxi}.
\eeq
Moreover, there should be a transformation that achieves this upper bound:
if the state is known, any number of its copies can be produced. As a result, a
 transformation which consists of the unambiguous state identification and 
the corresponding preparation achieves the bound of Eq. ({\ref{maxi}).
 However, the
more reasonable (and physically fruitful) transformation of~[2] has this
 maximal efficiency too.
The importance of Eq.~({\ref{maxi}) is 
 that  generalizations of this result~[20, 21] are valid for an arbitrary 
number of linearly independent states. For example, if $m$ copies of
three pure non-orthogonal states are used to produce the orthogonal ones,
the efficiency of the operation is bounded by~[20]
\beq
P({\rm success})\leq 1-\sum_{i=1}^3 k_i |[\phi^{\rm in}_1 \phi^{\rm in}_2
\phi^{\rm in}_3]|^{2m}/3,
\eeq
where coefficients $k_i$ depend on the relative orientation of the state 
vectors and $[\rm uvw]$ stands for the triple product of the vectors, i.e.,
the determinant of their components, in any basis.

\bigskip\noindent{\bf Acknowledgments}\medskip

Part of this work was performed during Elsag-Bailey-ISI research meeting on 
quantum
 computation, Turin 1998. It is a pleasure to thank Dagmar Bru\ss, Isaac
Chuang and Lev Levitin for useful discussions. The help of Chris Fuchs and 
Asher 
Peres is gratefully acknowledged. 
This work was supported by a grant from Technion Graduate School. 

\bigskip\noindent{\bf Appendix}\medskip
\renewcommand{\theequation}{A.\arabic{equation}}

\noindent Parameters of the cloning machine:
\setcounter{equation}{0}
\beq 
a=\frac{1}{\cos2\theta}[\cos\theta (P+Q\cos2\theta)-\sin\theta(P-Q\cos2\theta)]
, \eeq
\beq
b=\frac{1}{\cos2\theta}P\sin2\theta(\cos\theta-\sin\theta),
\eeq
\beq
c=\frac{1}{\cos2\theta}[\cos\theta (P-Q\cos2\theta)-\sin\theta(P+Q\cos2\theta)]
,\eeq
\beq
P=\frac{1}{2}\frac{\sqrt{1+\sin2\theta}}{\sqrt{1+\sin^2 2\theta}},
\eeq
\beq
Q=\frac{1}{2}\frac{\sqrt{1-\sin2\theta}}{\cos2\theta}.
\eeq
Probabilities of error for the output of cloning machine:
\beq
{\rm PE}(\theta)=\frac{1}{2}-\frac{1}{2}|\cos^2\theta-\sin^2\theta|,
\eeq
\beq
{\rm PE_d}(\theta)=\frac{1}{2}-\frac{1}{\sqrt{2}}
\frac{\sqrt{\cos^2 2\theta(13-10\cos4\theta+\cos8\theta+6\sin2 \theta-
2\sin6\theta)}}{\sqrt{(3-\cos4\theta)^3}}.
\eeq
Probabilities of error in more general case:
\begin{eqnarray}
{\rm PE}(\vartheta,\varphi) & = &
\frac{1}{2}-\frac{1}{2}|\cos^2\vartheta-\cos^2\varphi \sin^2\vartheta| 
\nonumber \\
 & & \times \sqrt{\cos^2\vartheta+\cos^2\varphi \sin^2\vartheta
+2 \sin^2\varphi\sin^2\vartheta+\cos\varphi\sin2\vartheta}
\end{eqnarray}

\beq
{\rm PE_d}(\vartheta,\varphi)= 
\frac{1}{2}-\frac{1}{2}|\cos^2\vartheta-\cos^2\varphi \sin^2\vartheta|
\sqrt{F(\vartheta, \varphi)},
\eeq
where $F$ is
\begin{eqnarray}
F(\vartheta, \varphi) & = &\cos^4\vartheta+\cos^2\vartheta\sin^2\vartheta 
(3-\cos2\varphi) \nonumber \\ 
& + &4\cos\varphi\cos\vartheta\sin^2\varphi\sin^3\vartheta+(5-\cos4\varphi)
\sin^4\vartheta/4. \nonumber
\end{eqnarray}

\newpage\frenchspacing 
\noindent{\bf References}
\medskip
\begin{enumerate}
\item W. K. Wootters and W. H. Zurek, Nature {\bf 299}, 802 (1982).
\item H. Bechman-Pasquinucci, B.  Huttner, and N. Gisin, Phys. Lett. A 
{\bf 242}, 198 (1998).
\item D. Bru{\ss}, D. P. DiVinzenzo, A. Ekert, C. A. Fuchs, C. Macchiavello,
J. A. Smolin, Phys. Rev. A {\bf 57}, 2368 (1998).  
\item A. Barenco, D.Deutch, A. Ekert, and R. Jozsa , Phys. Rev. Lett. 
{\bf 74}, 4083  (1995).
\item A. Peres, {\it Quantum Theory: Concepts and Methods\/}
(Kluwer, Dordrecht, 1993), Ch.~9.
\item  C. A. Fuchs, {\em Distinguishability and Accessible Information
 in Quantum Theory},
(Ph. D. thesis, University of New Mexico, 1995), e-print 
quant-ph/9601020.
\item E. B. Davies, {\it Quantum Theory of Open Systems\/} 
(Academic Press, New York, 1976), Ch.~1.
\item K. Kraus, {\it States, Effects and Operations\/}
(Springer-Verlag, Berlin, 1983).
\item M. Czachor and M. Kuna, Phys. Rev. A, {\bf 58}, 128 (1998), and 
references therein.
\item I. L. Chuang and M. A. Nielsen, J. Mod. Opt. {\bf 44}, 2455 (1997)
\item C. W. Helstrom, {\it Quantum Detection and Estimation Theory\/}
(Academic Press, New York, 1976).
\item C. A. Fuchs and C. M. Caves, Open Sys. Inf. Dyn. {\bf 3}, 1 (1995);
 C. A. Fuchs and J. van de Graaf, e-print quant-ph/9712042.
\item A. S. Holevo, {\it Coding Theorems for Quantum Channels\/}, in
{\em Tamagawa University Reserch Review} (1998), e-print quant-ph/9809023
\item G. M. D'Ariano and H. P. Yuen, Phys. Rev. Lett. {\bf 76}, 2832 
(1996). 
\item L. B. Levitin, Open Sys. Inf. Dyn. {\bf 2}, 319 (1994).
\item C. H. Bennett, D. P. DiVincenzo, J. S. Smolin and W. K. Wootters,
Phys. Rev. A {\bf 54}, 3824 (1996); V. Vedral and M. Plenio, Phys. Rev. A
{\bf 57}, 1619 (1998). 
\item C. A. Fuchs, private communications.
\item A. Peres, Phys. Rev. Lett. {\bf 77}, 1413 (1996); M. Horodecki,
P. Horodecki and R. Horodecki, Phys. Lett. A {\bf 223}, 1 (1996);
N. J. Cerf, C. Adami and R. M. Gingrich, e-print quant-ph/9710001.
\item D. Dieks, Phys. Lett. A {\bf 126}, 303 (1988); A. Peres, Phys. Lett. A
 {\bf 128}, 19 (1988).
\item A. Peres and D. R. Terno, J. Phys. A {\bf 31}, 7105 (1998).
\item M. Ban, K. Kurokawa, R. Monmose and O. Hirota, Int. J. Theor. Phys.
{\bf 36}, 1269 (1997); A. Chefles and S. M. Barnett, e-print quant-ph/9808018.
\end{enumerate}\vfill
\bigskip
\vfill\noindent CAPTIONS OF THE FIGURES\bigskip

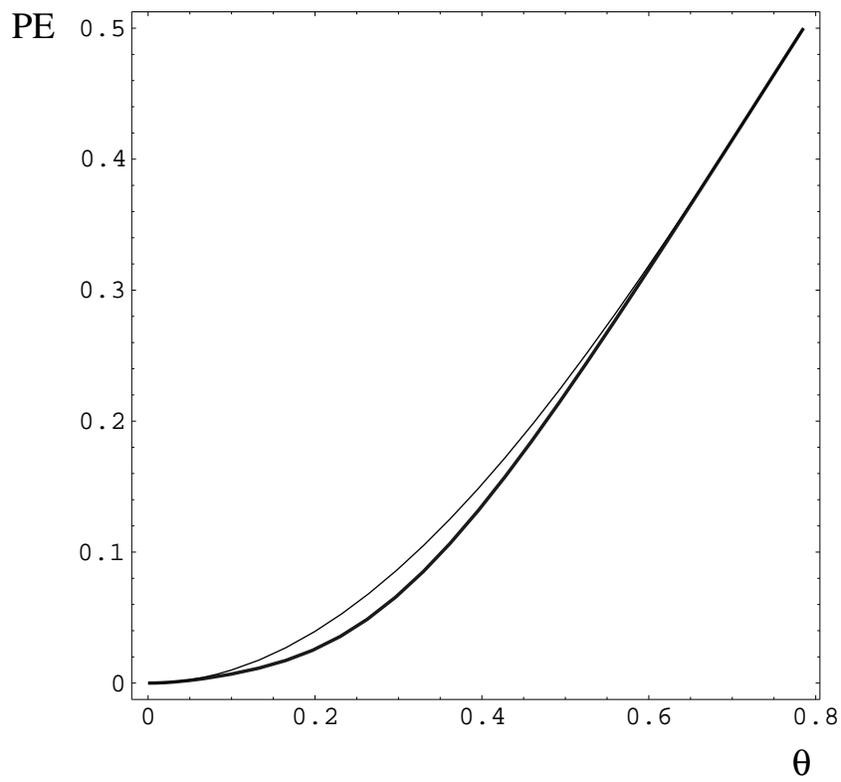
\noindent{\bf Figure 1.} Probabilities of Error:

\rule{15mm}{0.2mm}  PE in the
optimal measurement

\rule{15mm}{0.5 mm} PE in the optimal measurement after copying and 
disentanglement\medskip

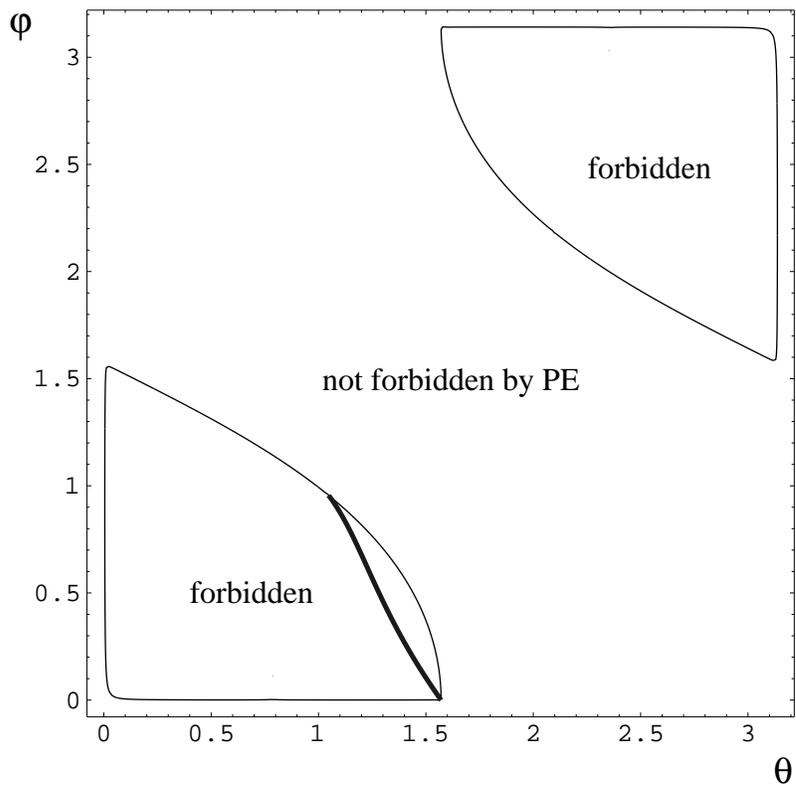
\noindent{\bf Figure 2.} Domains of the parameters for the 
disentanglement transformation according to the PE criterion

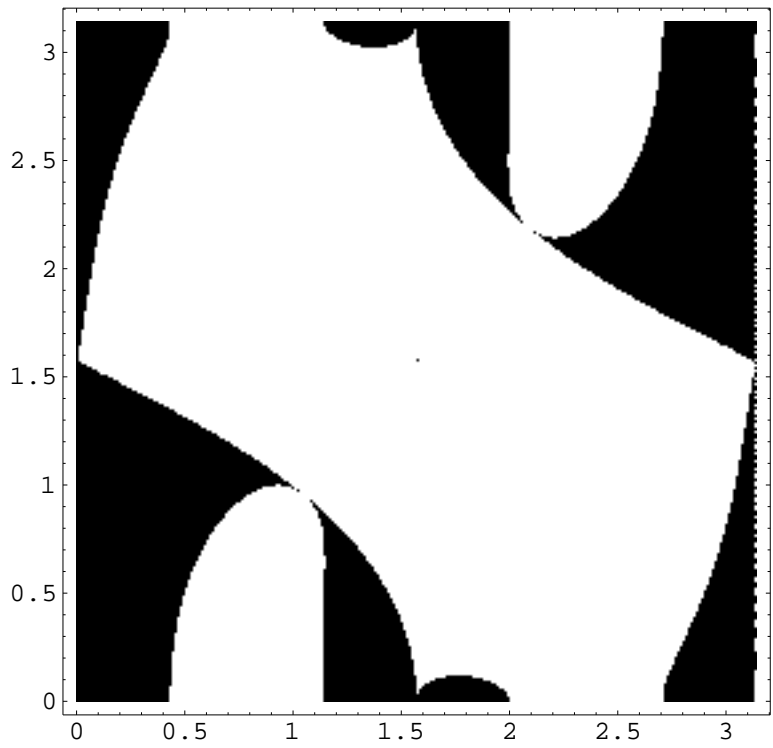
\noindent{\bf Figure 3.}  Domains of the parameters for the 
disentanglement transformation according to $\Delta S$ criterion

The forbidden domain is drawn in black.
\end{document}